\documentstyle[prl,aps,epsfig,twocolumn]{revtex}
\begin{document}
\draft
\flushbottom
\twocolumn[
\hsize\textwidth\columnwidth\hsize\csname @twocolumnfalse\endcsname

\title{Measuring billiard eigenfunctions with arbitrary trajectories} 
\author{Debabrata Biswas}
\address{
Theoretical Physics Division \\
Bhabha Atomic Research Centre \\
Mumbai 400 085, INDIA}
\date{\today}
\maketitle
\begin{abstract}
We propose a method of measuring approximate quantum eigenfunctions
in polygonalized billiard geometries, based on a quasiclassical
evolution operator having a (smoothened) Perron-Frobenius
kernel modulated by a phase arising from quantum considerations. 
We show that quasiclassical evolution
differs from semiclassical (or quantum) evolution but the operators have 
the same set of eigenfunctions. We demonstrate this by determining the
quasiclassical eigenfunctions of the polygonalized stadium billiard
using arbitrary trajectories and comparing this
with the exact quantum stadium eigenfunctions.
\end{abstract}

\pacs{PACS number(s): 05.45.Mt, 03.65.Sq, 31.15.Gy,05.45.Ac}
\date{today}
]
\narrowtext
\tightenlines

\newcommand{\be}{\begin{equation}}
\newcommand{\ee}{\end{equation}}
\newcommand{\bea}{\begin{eqnarray}}
\newcommand{\eea}{\end{eqnarray}}
\newcommand{\Lop}{{\cal L}}
\newcommand{\DB}[1]{\marginpar{\footnotesize DB: #1}}
\newcommand{\q}{\vec{q}}
\newcommand{\kt}{\tilde{k}}
\newcommand{\Lopn}{\tilde{\Lop}}

The quantum billiard problem consists of determining the eigenvalues and
eigenfunctions of the Helmholtz equation 
$\nabla^2 \psi(q) + k^2 \psi(q) = 0$
with $\psi(q) = 0$ on the billiard boundary $\partial B$ (Dirichlet
boundary condition). This simple wave equation 
arises in various contexts and has been used extensively to test
ideas of quantum chaos. It can describe acoustic
waves, modes in microwave cavities and has relevance
in studies on quantum dots where the motion of electrons can be regarded
as ``free'' inside an enclosure. The problem is analytically tractable
only for the small subset of ``integrable'' boundaries for which the
classical dynamics is regular. For other enclosures, 
the eigenstates must be computed numerically
and a number of efficient ``boundary'' methods exist that allow
us to study the eigenvalues and eigenfunctions.

Of particular interest is the determination of approximate quantum
eigenstates using classical quantities. 
While the old quantum theory of Bohr and co-workers works only for 
regular or integrable systems, 
modern semiclassical theories have respondend to the challenge 
posed by chaotic classical dynamics and the successful 
quantization of the Helium atom \cite{gt}
points to its success. The aim however is not necessarily linked to
the development of a cheap substitute for the computer intensive
numerical methods that determine the exact quantum states. While
this is a desirable consequence, semiclassical studies endevour to
provide an understanding of the quantum phenomenon in terms of
classical objects that we are so familiar with. Modern semiclassical
methods have indeed furthered our understanding of the 
quantum-classical correspondence. Thus, we are now aware of the
duality of quantum eigenenergies and classical periodic orbits -- 
a relationship that now forms the cornerstone of most semiclassical
theories \cite{gutzwiller}. The study of ``scars'' has also 
revealed the structure
of quantum eigenfunctions and classical trajectories have even
been used to construct semiclassical eigenfunctions of chaotic
systems \cite{semi_eigenfns}. Quantum states can thus be contemplated 
in classical
terms as a first approximation with corrections providing its
true quantum nature.

In this context, a  question that may be asked is the
following : {\em what is the degree of classical information
that is required in order to extract a first approximation of a 
quantum state} ? This is especially pertinent when the
system in question is chaotic or mixed with islands of
regularity interspersed in the chaotic sea. Since the quantum
state (or the quasiprobability distributions constructed
out of them) can essentially resolve phase space structures of the size
of a Plank cell,
is information finer than that redundant ? For billiards,
the de~Broglie wavelength $\lambda$, provides a relevant 
length scale that can be used effectively to probe
the boundary of the enclosure.  If a smooth billiard
boundary is polygonalized such that the short time
classical dynamics is well approximated,
the two billiards are semiclassically equivalent, provided, $\lambda$
is larger than the average length of edges of the
polygon \cite{db_00_1}. Thus, instead of the full chaotic dynamics
of the stadium billiard, one may as well consider the
dynamics of its polygonal counterpart for a given de~Broglie
wavelength. The polygonalization concept is implicit even 
in the numerical computation of the ``exact'' eigenvalues using
boundary methods in which the perimeter is discretized with
the number of points $N \sim {\cal L}/\lambda$ where ${\cal L}$ is
the perimeter of the enclosure. This idea has however not been
used to determine {\em semiclassical} eigenvalues or eigenfunctions.
Indeed, by most accounts a polygonalized approach to 
semiclassics is bound to be even more difficult since periodic
orbit quantization of polygons has proved largely unsuccessful 
\cite{shudo}. Also, it 
is generally believed that diffractive contributions must be
included even for obtaining a first approximation of a polygonal
quantum state.
On the other hand, it has recently been 
shown \cite{db_00_2} that closed alomost-periodic (CAP) orbits
also contribute in generic polygonal enclosures with weights
that are comparable to periodic orbits. Besides, they are more
numerous and hence indispensable for semiclassical quantization.
Thus, the failure of periodic orbit quantization in polygons
is not so much due to diffraction as due to the neglect of
CAP orbits. The modified periodic orbit theory for polygons
should in principle be able to provide approximate quantum
eigenvalues although CAP contributions can be difficult to
evaluate.

A recently developed time domain technique \cite{db_98,db_01} 
of determining quantum eigenvalues in marginally stable billiard
geometries makes use of arbitrary classical trajectories. 
The algorithm involves shooting trajectories in
various directions from a point interior to the billiard
(call it $q'$) and at each time
step, recording the fraction of trajectories
that are in an $\epsilon$ neighbourhood of a point $q$, weighted by
a phase arising from quantum considerations.
The peak positions in the power spectrum of this weighted fraction,
$F(t)$, are related to the quantum eigenvalues and as we
shall show in this letter, the heights of the peaks are a 
measure of the quantum
eigenfunctions at the point $q$. Thus, the arbitrary trajectory
quantization method (ATQM) is aptly suited for polygonalized
billiards and we shall demonstrate for the stadium billiard that
the quasiclassical eigenfunctions of the polygonalized
stadium do approximate the ``exact'' eigenfunctions of the 
full stadium.

The ATQM bypasses the need to evaluate periodic or CAP orbits
but is nevertheless based on the modified periodic orbit theory
for polygons. The parameter $\epsilon$ (${\cal O}(1/k)$) embodies
the contribition of CAP orbits and must be nonzero for any
calculation at finite $k$. In the following, we shall first
use a plane wave ansatz to show that the peak heights in the power
spectrum are a measure of the approximate quantum eigenfunction
and then demonstrate this for a polygonalized stadium.

We define the quasiclassical \cite{quasi} propagator 
$\Lop_{qc}^t(\varphi)$ restricted to the 
$\varphi$ invariant surface as

\be
\Lop_{qc}^t(\varphi)\circ \phi(q) = \int~dq'~\delta(q - q'^t(\varphi))
e^{-i\nu(t)\pi/2}~\phi(q')
\ee

\noindent
where $q'^t(\varphi)$ is the position at time $t$
of a trajectory which starts
at $q'$ ($t=0$) on an invariant surface labelled by $\varphi$
and the energy $E$.
The phase $\nu(t) = \nu(q'^t(\varphi))$ depends on the caustic structure
of the trajectory and is identical to the phase in
the semiclassical propagator \cite{gutzwiller}. 
When $\nu(t)$ is identially zero (as in
case of Neumann boundary conditions), $\Lop_{qc}^t(\varphi)$ reduces
to the Perron-Frobenius operator on the $\varphi$ invariant surface.
Note that rational polygonal billiards
have a second constant of motion
due to which the invariant surface is 2-dimensional \cite{varphi}. 
On successive reflections,
the trajectory changes its momentum following the laws of reflection but
continues to live on the same invariant surface labelled by $E$ and $\varphi$.
Thus, $\varphi$ also labels distinct trajectories that
start from a given point $q'$. 

The full quasiclassical evolution operator is defined as

\bea
\nonumber
\Lop^t_{qc} \circ \phi(q) & = & \int~d\varphi~\Lop^t_{qc}(\varphi) \\ \nonumber
& = & \int~dq' \left \{ \int~d\varphi~\delta(q - q'^t(\varphi))
e^{-i\nu(t)\pi/2} \right \}\phi(q') \\
& = & \int~dq'~K_{qc}(q,q',t)~\phi(q')
\eea

\noindent
Note that quasiclassical evolution differs from semiclassical
evolution. To illustrate this, 
consider a particle in a one-dimensional box (Dirichlet boundary 
conditions) and consider the evolution
of the quantum eigenfunction $\psi_n(q) = e^{ik_n q} - e^{-ik_n q}$, 
$k_n = n\pi/L$ . Its time
evolution in quantum mechanics is simply $e^{-iE_n t} \psi(q)$ where
$E_n = \hbar^2 k_n^2/2m$. Its quasiclassical evolution, 
$\Lop^t_{qc}(+) \circ \psi_n(q)$, is given by

\be
\left ( e^{ik_n q^{-t}(+v)} - 
e^{-ik_n q^{-t}(+v)}
\right ) e^{-i\pi n(q^{-t}(+v))}
\ee

\noindent
where $n(q^{-t}(+v))$ is the number
of reflections suffered by a trajectory in time $-t$ with initial position $q$
and initial velocity $+v$. Similarly, $\Lop^t_{qc}(-) \circ \psi_n(q)$ is

\be
\left ( e^{ik_n (q^{-t}(-v))} - 
e^{-ik_n(q^{-t}(-v))}
\right ) e^{-i\pi n(q^{-t}(-v))}
\ee

\noindent
with the $-$ sign in $\Lop^t_{qc}(-)$ denoting negative velocity.
Note that the flow is such that the velocity changes sign at every 
reflection from the walls at $q=0$ and $q=L$ while $n(t)$ increments
by one at each of these instants. For the flow $q^{-t}(+v)$, the
reflections occur at $t_n^+ = (q + nL)/v$ so that for $t_0^+ < t < t_1^+$,
$q^{-t}(+v) = v(t-t_0^+) = vt - q$. Similarly, for the flow $q^{-t}(-v)$,
the reflections occur at $t_n^- = (L-q + nL)/v$ and for 
$t_0^+ < t < t_1^+$, $q^{-t}(-v) = L - v(t-t_0^-) = 2L - vt - q$.
It follows hence that

\be
\Lop_{qc}^t(\pm)\circ \psi_n(q) = e^{ik_n(q~\mp~vt)} - e^{-ik_n(q~\mp~vt)}
\ee

\noindent
for all $t$. Thus 

\be
(\Lop^t_{qc}(+) + \Lop^t_{qc}(-) ) \circ \psi_n(q) = 2\cos(k_n v t) \psi_n(q)
\ee

\noindent
In other words, $\psi_n(q)$ is also an eigenfunction of the full quasiclassical
evolution operator $\Lop^t_{qc} = \Lop^t_{qc}(+) + \Lop^t_{qc}(-)$.
Note that in the Neumann case, $\psi_n(q)$ is not an eigenfunction; rather
$e^{ik_nq} + e^{-ik_nq}$ is an eigenfunction with $n(t) = 0$.

For a general 2-dimensional billiard, there is strong numerical evidence
to suggest that a (real) plane wave superposition can be used to 
construct eigenfunctions \cite{heller,ford_et_al,db_ss}. We shall adopt
here the view that a finite plane wave superposition does yield at least 
good approximate eigenfunctions. For polygonalized billiards, 
the semiclassical wavefunction can be expressed as 

\be
\psi(q) = \sum_{j=1}^{M} A_j e^{ik\cos(\mu_j)x + ik\sin(\mu_j)y}.
\label{eq:plane_waves}
\ee

\noindent
where $A_j$ are constants \cite{keller_60} and the number
of terms $M$ in the expansion is determined by closure
of the wave vector $\vec{k} = (k\cos \mu_j, k\sin \mu_j)$ 
under reflection from the edges.

For this finite superposition of plane waves, 
the boundary condition 
$\psi(q) = 0$ on $\partial B$ can be satisfied if the waves
vanish in pairs with an incident wave giving rise to a reflected
wave. Thus on the $l$th segment $y = a_lx + b_l$, we must have 

\bea \nonumber 
& \; & A_j e^{i(k \cos\mu_j + a_l k \sin\mu_j)x + ib_lk\sin\mu_j} \\
& +  & A_{j'} e^{i(k \cos\mu_{j'} + a_l k 
\sin\mu_{j'})x + ib_lk\sin\mu_{j'}} = 0 .  \label{eq:bc1}
\eea  

\noindent
Assuming that $\mu_{j'}$ is related to $\mu_j$ through the laws of
reflection, it is easy to show that 

\be
 \cos\mu_j + a_l \sin\mu_j  =   \cos\mu_{j'} + a_l \sin\mu_{j'}
\ee

\noindent
Thus, eq.~(\ref{eq:bc1}) reduces to

\be
A_j e^{ib_l k\sin\mu_j} + A_{j'} e^{ib_l k\sin\mu_{j'}} = 0
\label{eq:cond1}
\ee

\noindent
where $\mu_{j'} = \pi - \mu_j + 2\theta_l$ and $\theta_l$ is the angle
between the positive $X$-axis and the outward normal to the $l$th line
segment. 

Note that for each of the $K$ segments on the boundary, the $j$th wave
has in general a different reflected wave as a counterpart so that 
eq.~(\ref{eq:cond1}) gives $K$ different expressions for $A_j$.
In general (barring exceptions such as the rectangle billiard),
these ``boundary conditions'' can be satisfied only approximately
as we shall argue below.
Recall that for the numerical determination of exact eigenvalues
using a plane wave basis, the boundary is discretized ($N$ points)
and an appropriate measure (such as a determinant) is used 
to determine the eigenstates which satisfy the boundary condition
at these points. Convergence can be achieved by increasing $N$
so that as $N \rightarrow \infty$, the boundary condition
is satisfied exactly. In contrast, 
the number of terms in eq.~\ref{eq:plane_waves} is fixed.
Thus, if the exact eigenfunction contains additional plane waves,
the boundary condition will be satisfied {\em approximately}
and the plane wave expansion of  eq.~\ref{eq:plane_waves} can
only give an approximate quantum eigenfunction.

We shall now establish
that the (finite) plane wave superposition (eq.~\ref{eq:plane_waves})
is {\em also} an approximate eigenfunction 
of the quasiclassical evolution operator provided the set of
``quantization'' conditions given by eq.~(\ref{eq:cond1}) are satisfied.

Consider therefore the plane wave superposition of 
eq.~(\ref{eq:plane_waves}).
Its quasiclassical evolution is given by 

\be
\Lop^t_{qc}(\varphi) = \sum_{j=1}^M A_j e^{ik_x x^{-t}(\varphi) +
ik_y y^{-t}(\varphi)} e^{-in(t)}
\ee

\noindent
where $k_x = k\cos(\mu_j)$, $k_y = k\sin(\mu_j)$
while $x^{-t}(\varphi)$ and $y^{-t}(\varphi)$ denote the flow at time
$-t$ with initial position ($x,y$) and velocity 
($v\cos\varphi,v\sin\varphi$). For short times, this is given by

\be \nonumber
\Lop^t_{qc}(\varphi) = \sum_{j=1}^M A_j e^{ik_x (x - v \cos\varphi t)
+ ik_y (y - v \sin\varphi t) } .
\ee

\noindent
As before, we shall first determine the evolution of a single
wave after reflection from one of the segments, $y = a_lx + b_l$.
For the flow, ($x^{-t}(\varphi),y^{-t}(\varphi)$), reflection from 
the line segment takes place at $t_0 = (x - x_0)/(v\cos\varphi) =
(y - y_0)/(v\sin\varphi)$ where ($x_0,y_0$) is the point of impact.
The flow at a time $t$ after the reflection is given by

\bea
\nonumber
x^{-t}(\varphi) & = & x(t) =  x_0 + v\cos(\varphi - 2\theta_l)(t-t_0) \\
y^{-t}(\varphi) & = & y(t) =  y_0 + v\sin(\varphi - 2\theta_l)(t-t_0) .
\eea

\noindent
It is easy to verify that after one reflection from the segment
$y= a_l x + b_l$, the wave
$A_j e^{ik\cos\mu_j x + k\sin\mu_j y}$ evolves quasiclassically to

\be
 A_{j'}  e^{k\cos\mu_{j'} (x - v\cos\varphi t) + k\sin\mu_{j'}
 (y - v\sin\varphi t)}
\ee

\noindent
where $\mu_{j'} = \pi - \mu_j + 2\theta_l$ and

\be
A_j e^{ikb_l \sin\mu_j} + A_{j'} e^{ikb_l \sin\mu_{j'}} = 0
\label{eq:cond_qc}
\ee

\noindent
Thus, after one reflection, the finite plane wave superposition
assumes the form for small $t$
provided the reflected waves are included in the superposition
and the ``quantization conditions'' are (approximately) 
satisfied for a given value of $k_n$. 
It follows that $\psi_n(q)$ is an (approximate) eigenfunction 
of $\Lop_{qc} = \int~d\varphi~\Lop_{qc}(\varphi)$ :

\bea
\Lop^t_{qc} \circ \psi_n(q) & = & \int~d\varphi~\Lop_{qc}(\varphi) \psi_n(q) \\
& = & \int~d\varphi~e^{-ik_nvt\cos(\varphi - \mu_j)} 
  \sum_j A_j e^{iS_j(k_n)} \\
& = & 2\pi J_0(k_nvt)~\psi_n(q)
\eea

\noindent
where $S_j(k_n) = k_n\cos(\mu_j) x + k_n\sin(\mu_j) y$.
Thus, {\em a finite plane wave superposition can be an
approximate semiclassical 
and a quasiclassical eigenfunction under identical conditions}. 

We shall now demonstrate our result for a stadium billiard consisting
of two parallel straight segments of length $2$ joined on either end 
by a semicircle of unit radius. For the evaluation of the quasiclassical
eigenfunctions, we shall consider a polygonalized enclosure where
each semicircle is replaced by $12$ straight edges of equal length.
In order to determine the quasiclassical eigenfunctions, we shall 
first evaluate a smoothened quasiclassical kernel

\bea \nonumber
K_{qc}(q,q',t) & = & \int~d\varphi~\delta_\epsilon(q - {q'}^t(\varphi))
e^{-i\pi n({q'}^t(\varphi))} \\ & = & \sum_n \psi_n(q) {\psi_n}^*(q')
\Lambda_n(t) \label{eq:smoothen}
\eea

\noindent
as a function of time. Thus is
achieved by shooting trajectories from a point $q'$ at various angles
and evaluating the fraction of trajectories in a cell of size $\epsilon$ 
\cite{other_delta} at $q$,
weighted by the phase $e^{-i\pi n({q'}^t(\varphi))}$.
Since $\Lambda_n = 2\pi J_0(k_nvt)$, for $v=1$, a fourier
transform of $K_{qc}(q,q',t)$  has peaks at $k = k_n$ and the
heights are proportional to $\psi_n(q)$. 

Note that the smoothening
of the delta function kernel is essential in order to accomodate
closed almost-periodic orbits and shows up naturally in an 
alternate proof (of the identity of the semiclassical and quasiclassical
eigenvalues) involving the trace of the quasiclassical and 
semiclassical propagators \cite{db_01}. In the present formalism involving
plane waves, smearing of the kernel 
leads to a modified ``quantization condition'' for the quasiclassical
eigenfunctions. We shall however
ignore these complications and merely reiterate that the 
parameter $\epsilon \sim 1/k$.

\begin{figure}[tbp]
\hspace*{0.5cm}\epsfig{figure=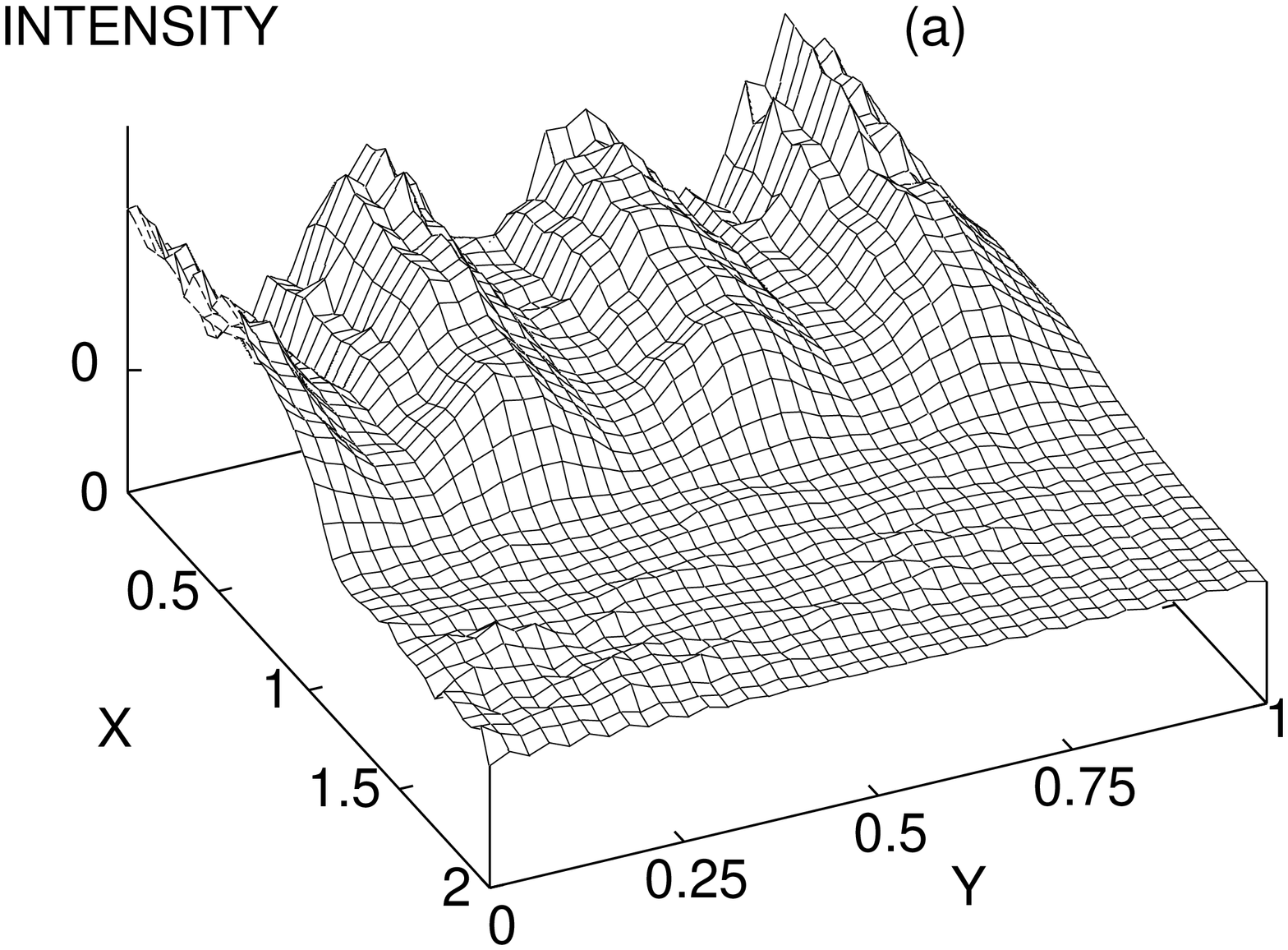,width=5.cm,angle=270}
\hspace*{0.5cm}\epsfig{figure=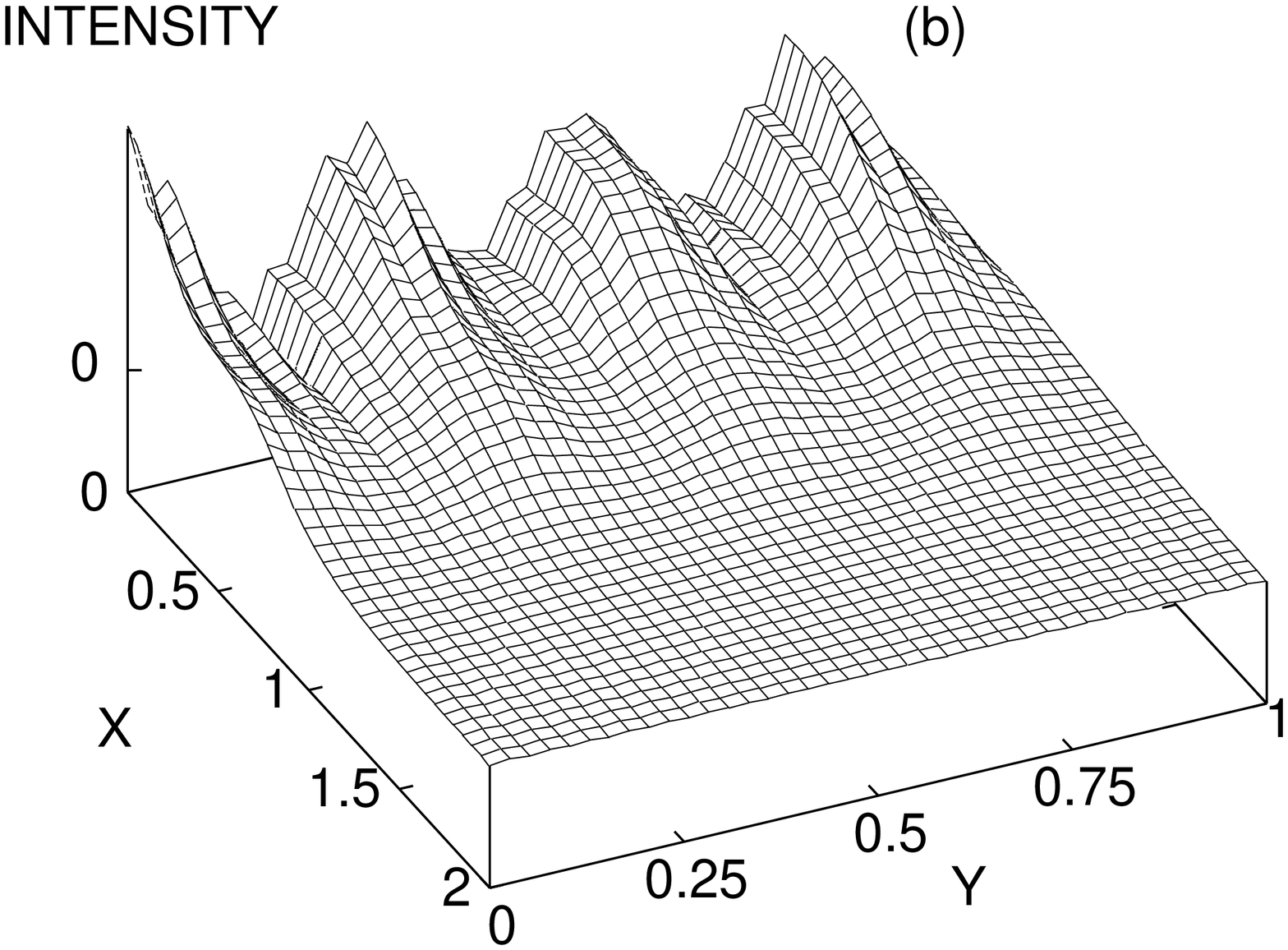,width=5.cm,angle=270}
\caption[ty]{(a) A quasiclassical bouncing ball eigenfunction of 
the polygonalized stadium 
and (b) its quantum counterpart in the smooth stadium}
\label{fig:1}
\end{figure}

Fig~\ref{fig:1}a shows a ``bouncing ball''
quasiclassical eigenfunction 
intensity $|\psi_n(q)|^2$ at $k=10.97$ in the quarter stadium.
 
\begin{figure}[tbp]
\hspace*{0.5cm}\epsfig{figure=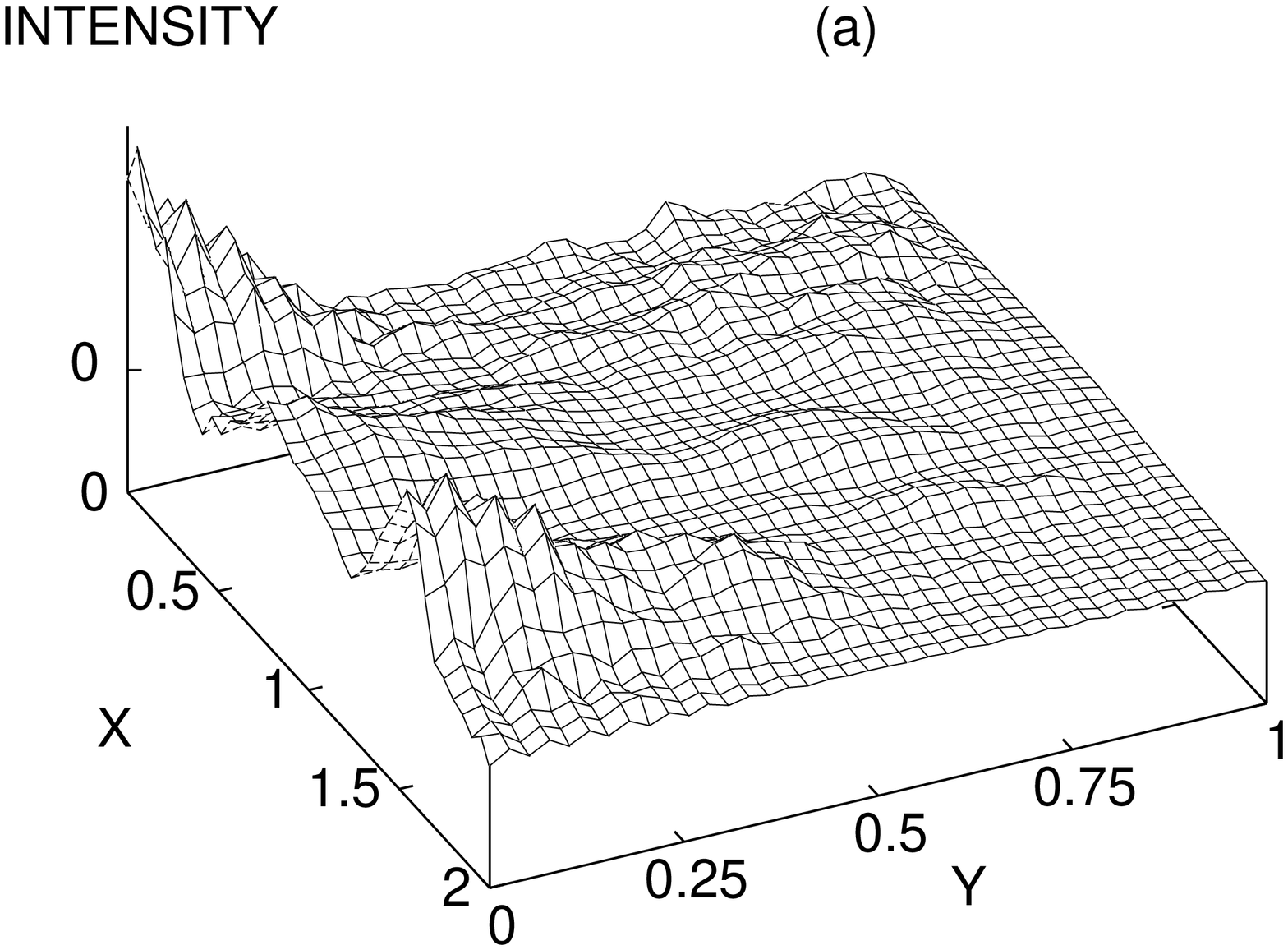,width=5.cm,angle=270}
{\vspace*{0.02in}}
\hspace*{0.5cm}\epsfig{figure=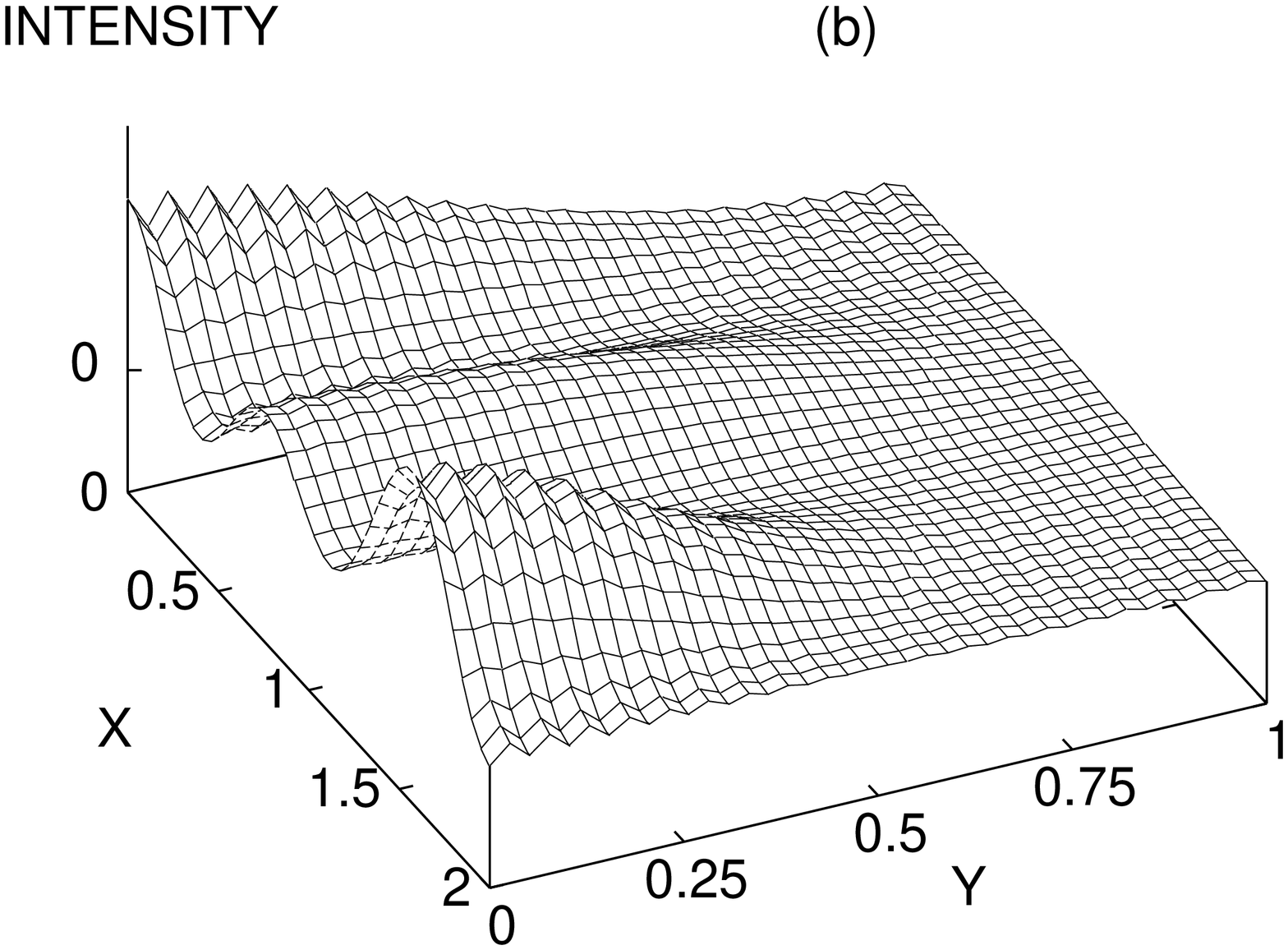,width=5.cm,angle=270}
\caption[ty]{(a) A quasiclassical eigenfunction peaked along
the X axis and (b) its quantum counterpart}
\label{fig:2}
\end{figure}

\noindent
while the corresponding quantum eigenfunction in the 
stadium at $k=11.05$ is shown in fig.\ref{fig:1}b. 
An example of a quasiclassical eigenfunction at $k=4.02$ peaked along 
the X-axis is shown in fig.~\ref{fig:2}a along with its
quantum conterpart at $k=4.38$.
The quasiclassical eigenfunction clearly provides a first approximation
of the quantum eigenfunction in both cases. Eigenfunctions of
other billiards including triangles have also been  obtained.
Details of this work will be published elsewhere.

In conclusion, we have demonstrated that the quasiclassical
eigenfunctions determined using arbitrary trajectories in a 
polygonalized chaotic enclosure, approximates
the quantum eigenfunctions of the smooth billiard.
We have also shown that a finite plane wave expansion is an 
approximate eigenfunction
of the quantum and quasiclassical evolution operators under identical
conditions.

\newcommand{\PR}[1]{{Phys.\ Rep.}\/ {\bf #1}}
\newcommand{\PRL}[1]{{Phys.\ Rev.\ Lett.}\/ {\bf #1}}
\newcommand{\PRA}[1]{{Phys.\ Rev.\ A}\/ {\bf #1}}
\newcommand{\PRB}[1]{{Phys.\ Rev.\ B}\/ {\bf #1}}
\newcommand{\PRD}[1]{{Phys.\ Rev.\ D}\/ {\bf #1}}
\newcommand{\PRE}[1]{{Phys.\ Rev.\ E}\/ {\bf #1}}
\newcommand{\JPA}[1]{{J.\ Phys.\ A}\/ {\bf #1}}
\newcommand{\JPB}[1]{{J.\ Phys.\ B}\/ {\bf #1}}
\newcommand{\JCP}[1]{{J.\ Chem.\ Phys.}\/ {\bf #1}}
\newcommand{\JPC}[1]{{J.\ Phys.\ Chem.}\/ {\bf #1}}
\newcommand{\JMP}[1]{{J.\ Math.\ Phys.}\/ {\bf #1}}
\newcommand{\JSP}[1]{{J.\ Stat.\ Phys.}\/ {\bf #1}}
\newcommand{\AP}[1]{{Ann.\ Phys.}\/ {\bf #1}}
\newcommand{\PLB}[1]{{Phys.\ Lett.\ B}\/ {\bf #1}}
\newcommand{\PLA}[1]{{Phys.\ Lett.\ A}\/ {\bf #1}}
\newcommand{\PD}[1]{{Physica D}\/ {\bf #1}}
\newcommand{\NPB}[1]{{Nucl.\ Phys.\ B}\/ {\bf #1}}
\newcommand{\INCB}[1]{{Il Nuov.\ Cim.\ B}\/ {\bf #1}}
\newcommand{\JETP}[1]{{Sov.\ Phys.\ JETP}\/ {\bf #1}}
\newcommand{\JETPL}[1]{{JETP Lett.\ }\/ {\bf #1}}
\newcommand{\RMS}[1]{{Russ.\ Math.\ Surv.}\/ {\bf #1}}
\newcommand{\USSR}[1]{{Math.\ USSR.\ Sb.}\/ {\bf #1}}
\newcommand{\PST}[1]{{Phys.\ Scripta T}\/ {\bf #1}}
\newcommand{\CM}[1]{{Cont.\ Math.}\/ {\bf #1}}
\newcommand{\JMPA}[1]{{J.\ Math.\ Pure Appl.}\/ {\bf #1}}
\newcommand{\CMP}[1]{{Comm.\ Math.\ Phys.}\/ {\bf #1}}
\newcommand{\PRS}[1]{{Proc.\ R.\ Soc. Lond.\ A}\/ {\bf #1}}


\begin{references}
\bibitem{quasi} We reserve the term ``semiclassical'' for those
studies which involve an approximation
of the quantum propagator in the $\hbar \rightarrow 0$ limit
and use the term ``quasiclassical'' for the modified classical
propagator.
\bibitem{varphi} For rational polygons, the angle $\varphi$ measures
the linear momentum ($p\cos\varphi,p\sin\varphi$). It is not directly
a conserved quantity but serves to label the invariant surface
which consists of a finite number of copies of the polygon glued
appropriately along the edges.
\bibitem{other_delta} Several other smoothened kernels can 
be used - see \cite{db_01} for details.
\end{references}
\end{document}